# USING DATA MINING TECHNIQUES FOR DIAGNOSIS AND PROGNOSIS OF CANCER DISEASE

Shweta Kharya

Sr.Assistant Professor,
Bhilai Institute of Technology,
Durg-491 001, Chhattisgarh, India
Shweta.bitdurg@gmail.com

## ABSTRACT

*Breast cancer is one of the leading cancers for women in developed countries including India. It is the second most common cause of cancer death in women. The high incidence of breast cancer in women has increased significantly in the last years. In this paper we have discussed various data mining approaches that have been utilized for breast cancer diagnosis and prognosis. Breast Cancer Diagnosis is distinguishing of benign from malignant breast lumps and Breast Cancer Prognosis predicts when Breast Cancer is to recur in patients that have had their cancers excised. This study paper summarizes various review and technical articles on breast cancer diagnosis and prognosis also we focus on current research being carried out using the data mining techniques to enhance the breast cancer diagnosis and prognosis.*

## 1. INTRODUCTION

Data mining is an essential step in the process of knowledge discovery in databases in which intelligent methods are applied in order to extract patterns. Breast cancer has become the primary reason of death in women in developed countries. The most effective way to reduce breast cancer deaths is to detect it earlier. Early diagnosis needs an accurate and reliable diagnosis procedure that can be used by physicians to distinguish benign breast tumors from malignant ones without going for surgical biopsy. The objective of these predictions is to assign patients to one of the two group either a *"benign"* that is noncancerous or a *"malignant"* that is cancerous. The prognosis problem is the long-term care for the disease for patients whose cancer has been surgically removed.

Predicting the outcome of a disease is one of the most interesting and challenging tasks where to develop data mining applications. The use of computers with automated tools, large volumes of medical data are being collected and made available to the medical research groups. As a result, data mining techniques has become a popular research tool for medical researchers to identify and exploit patterns and relationships among large number of variables, and made them able to predict the outcome of a disease using the historical datasets. The objective of this study is to summarize various review and technical articles on diagnosis and prognosis of breast cancer. In this paper we provided an overview of the current research being carried out on various breast cancer datasets using the data mining techniques to enhance the breast cancer diagnosis and prognosis.

## 2. AN OVERVIEW OF BREAST CANCER

Breast cancer is the most common cancer among Women. The malignant tumor develops when cells in the breast tissue divide and grow without the normal controls on cell death and cell division. Hence, cancer on breast tissue is called breast cancer. Worldwide, it is the most common form of cancer in females that is affecting approximately 10% of all women at some stage of their life. Although scientists do not know the exact causes of most breast cancer, they do know some of the risk factors that increase the likelihood of a woman developing breast cancer. These factors include such attributes as age, genetic risk and family history. Although breast cancer is the second leading cause of cancer death in women, the survival rate is high. With early diagnosis, 97% of women survive for 5 years or more years. The two main type's whereas chemotherapy and hormone therapy are systematic therapies.

## 3. METHODOLOGY

The main methodology used for this paper was through the survey of journals and publications in the field of medicine, computer science and engineering. The research focused on more recent publications.

## 4. RESEARCH FINDINGS

**4.1 Decision Trees** treatments for breast cancer are, local and systematic. Surgery and radiation are local treatments

In [1] the authors have explored the applicability of decision trees to do find a group with high-susceptibility of suffering from breast cancer. The goal was to find one or more leaves with a high percentage of cases and small percentage of controls. A case-control study was performed, composed of 164 controls and 94 cases with 32 SNPs available from the BRCA1, BRCA2 and TP53 genes. The data consists of information about tobacco and alcohol consumption. To statistically validate the association found, permutation tests were used. It has been found that a high-risk breast cancer group composed of 13 cases and only 1 control. These results show that it is possible to find statistically significant associations with breast cancer by deriving a decision tree and selecting the best leaf. A dataset collected by the Department of Genetics of the Faculty of Medical Sciences of Universidad Nova de Lisboa with 164 controls and 94 cases, all of them being Portuguese Caucasians. Of the 94 cases, 50 of them had its tumour detected after menopause in women above 60 years old, while the other 44 had its tumour detected before menopause, in women under 50 years old. The tumour type is ductal carcinoma (invasive and in situ). SNPs were selected with Minor Allele Frequency above or equal to 5% for european caucasian population (HapMap CEU). Tag SNPs were selected with a correlation coefficient r2 =

0:8. A total of 32 SNPs are available, 7 from the BRCA1 gene (rs16942, rs4986850, rs799923, rs3737559, rs8176091, rs8176199 and rs817619), 19 from the BRCA2 gene (rs1801406, rs543304, rs144848, rs28897729, rs28897758, rs15869, rs11571836, rs1799943, rs206118, rs2126042, rs542551, rs206079, rs11571590, rs206119, rs9562605, rs11571686, rs11571789, rs2238163 and rs1012130) and 6 from the TP53 gene (rs1042522, rs8064946, rs8079544, rs12602273, rs12951053 and rs1625895). SNPs belong to several parts of the gene: regulatory region, coding region or non-coding region. The genotyping was done with real time PCR (Taqman echnology). Tobacco and alcohol consumption were also used as attributes for the analysis.

Decision Tree Learning is one of the most widely used and practical methods for classification. In this method, learned trees can be represented as a set of if-then rules that improve human readability. Decision trees are very simple to understand and interpret by domain experts. A decision tree consists of nodes that have exactly one incoming edge, except the root node that has no incoming edges. A node with outgoing edges is an internal node, while the other nodes are called leaves or terminal nodes or decision nodes. The experiments,is conducted using Weka J48, C4.5 decision tree is generated . Several parameters were tested such as the confidence factor 6 O. used for pruning, whether to use binary splits or not, whether to prune the tree or not and the minimum number of instances per leaf. For each different combination of parameters, the average classification accuracy of the 10 folds is saved. The best combination of parameters is selected, with higher average classification accuracy on 10-fold cross validation. The final model is shown in Figure 1.

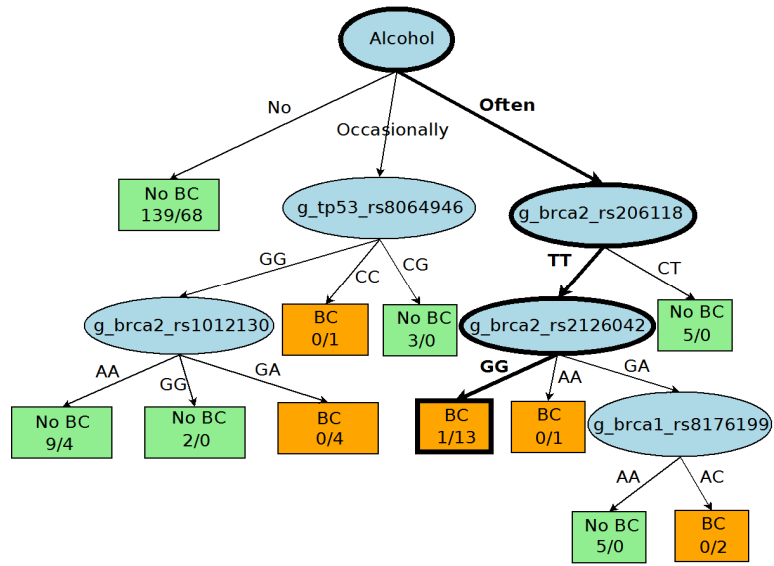

Figure 1. Decision Tree Model

With this methodology, the authors have showed that it is possible to find statistically significant associations from a breast cancer data set. However, this methodology needs to be evaluated in a larger set of examples in order to find associations with a higher degree of statistical confidence. Using a larger data set will also enable us to find correlations between a bigger set of genes and SNPs.

## 4.2 Digital Mammography Classification using Association Rule Mining and ANN

In [2] the authors have performed some experiments for tumor detection in digital mammography. In this paper different data mining techniques, neural networks and association rule mining, have been used for anomaly detection and classification. From the experimental results it is clear that the two approaches performed well, obtaining a classification accuracy reaching over 70% percent for both techniques. The experiments conducted, demonstrate the use and effectiveness of association rule mining in image categorization.

The real medical images used in the experiments were taken from the Mammographic Image Analysis Society (MIAS). It consists of 322 images, corresponding to three categories: normal, benign and malign. There were 208 normal images, 63 benign and 51 malign, which are considered abnormal. The abnormal cases are further divided in six categories: micro calcification, circumscribed masses, speculated masses, ill-defined masses, architectural distortion and asymmetry.

Figure 1 shows an overview of the categorization process adopted for both systems. The first step is represented by the image acquisition and image enhancement, followed by feature extraction. The last one is the classification.

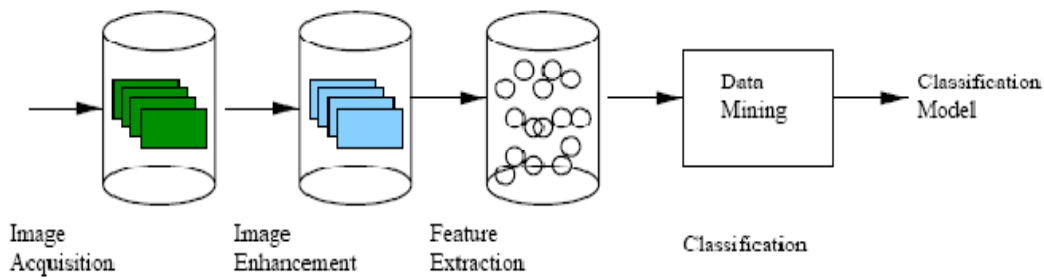

Figure 2. Image categorization process.

Mammograms are the images but difficult to interpret, and a preprocessing phase of the images is necessary to improve the quality of the images and make the feature extraction phase more reliable. Two Image Enhancement techniques: a cropping operation and an image enhancement has been performed before feature extraction. After cropping and enhancing of the images, which basically represents the data cleaning phase, features relevant to the classification are extracted from the cleaned images.

The existing features are:

    i.    The type of the tissue (dense, fatty and fatty glandular);
   ii.    The position of the breast: either left or right.

The above extracted features are computed over smaller windows of the original image. The original image is split in four parts as is shown in Figure 3 . These four regions are again splitted in other four parts to obtain more accurate extraction of the features and for the further investigation of the localization. For the sixteen sub-parts of the original image the statistical parameters were computed.

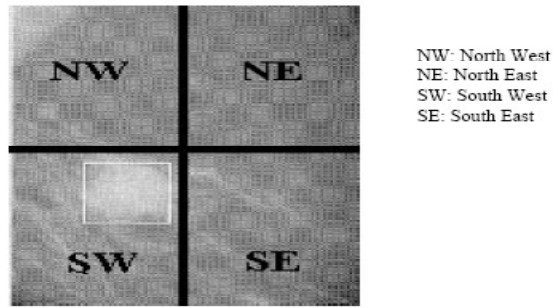

FIGURE-3

The four regions of the first division, and then, for each of the areas is further divided in four. To automatically categorize medical images on real mammograms two data mining techniques, association rule mining and neural networks is used.

### 4.2.1 Association rule based Classifier

Association rule mining aims at discovering associations between items in a transactional database. Given a set of transactions $D = \{T1,....., Tn\}$ and a set of items $I = \{i1, ...., im\}$ such that any transaction $T$ in $D$ is a set of items in $I$, an association rule is an implication of the form $A \Rightarrow B$ where the antecedent $A$ and the consequent $B$ are subsets of a transaction $T$ in $D$, and $A$ and $B$ have no common items. For the association rule to be acceptable, the conditional probability of B given A has to be higher than a threshold called minimum confidence. Association rules mining is a two-step process, in the first step frequent item-sets are generated (i.e. item-sets whose support is no less than a minimum support) and in the second step association rules are derived from the frequent item-sets obtained in the first step.

The apriori algorithm is used in order to discover association rules among the features extracted. After all the features are merged and put in the transactional database, the next step is applying the apriori algorithm for finding the association rules in the database constrained as described above with the antecedent being the features and the consequent being the category. Once the association rules are found, they are used to construct a classification system that categorizes the mammograms as normal, malign or benign. The most delicate part of the classification with association rule mining is the construction of the classifier itself. In the training phase, the apriori algorithm was applied on the training data to extract association rules. The support was set to 10% and the confidence to 0%. The success rate for association rule classifier was 69.11% on average. The results for the ten splits of the database are presented in Table 1.

| Database split | Success ration (percentage) |
|----------------|------------------------------|
| 1 | 67.647 |
| 2 | 79.412 |
| 3 | 67.647 |
| 4 | 61.765 |
| 5 | 64.706 |
| 6 | 64.706 |
| 7 | 64.706 |
| 8 | 64.706 |
| 9 | 67.647 |
| 10 | 88.235 |
| | Average: 69.11 |

Table 1 .Success ratios of association rule based classifier using 10 splits.

### 4.2.3 Neural Network based classifier system

The  neural network architecture consists of three layers: an input layer, a hidden one and an output layer. The number of nodes in the input layer represents the number of elements existing in one transaction in the database. In this experiment the input layer had 69 nodes. For the hidden layer, 10 nodes, while the output layer was consisting of only one node. The node of the output layer gives the classification for the image. It classifies the normal or abnormal cases.

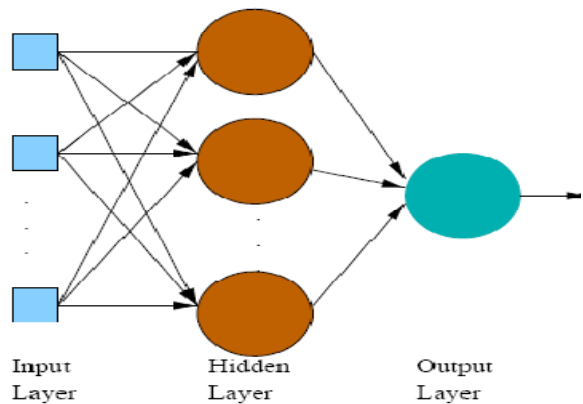

Figure 4. A 2-layer neural network.

During   training, the internal weights of the neural network are adjusted according to the transactions used in the learning process. For each training transaction the neural network receives the expected output. This allows the modification of the weights. In the next step, the trained neural network is used to classify new images.

| Database split | Success ration (percentage) |
|:---:|:---:|
| 1 | 96.870 |
| 2 | 90.620 |
| 3 | 90.620 |
| 4 | 78.125 |
| 5 | 81.250 |
| 6 | 84.375 |
| 7 | 65.625 |
| 8 | 75.000 |
| 9 | 56.250 |
| 10 | 93.750 |
| | Average: 81.248 |

Table 2. Success ratios of neural network based classifier for the 10 splits.

In [3] to classify the medical data set a neural network approach is adopted. The neural network is trained with breast cancer data base by using feed forward neural network model and back propagation learning algorithm with momentum and variable learning rate. The performance of the network is evaluated. The experimental result shows that by applying parallel approach in neural network model yields efficient result.

Neural Networks Parallelization Strategies as the Neural Networks are inherently parallel in nature, this technique is considered in this study to implement parallelism for calculating the output at each node in different layers of the network. Every neuron operates independently, processing the input receives, adjusting weights, and propagating its computed output thus a neuron is a natural level of parallelization for neural networks. Every neuron is treated as a parallel process.

Keeping in view of the significant characteristics of NN and its advantages for the implementation of the classification problem, Neural Network technique is considered for the classification of data related to medical field in this study. In this experiment the medical data related to breast cancer is considered. This database was obtained from the university of Wisconsin hospital, Madison from Dr. William H. Wolberg. This is publicly available dataset in the Internet.

Descriptions of Database:

- Number of instances 699
- Number of attributes: 10 plus the class attribute
- Attributes 2 through 10 will be used to represent instances
- Each instance has one of 2 possible classes: benign or malignant
- Class distribution: Benign : 458 (65.5%) ,Malignant : 241 (34.5%)

Following set of attributes have been used.

| Attribute | Domain |
|---|---|
| 1. Sample code number | id number |
| 2. Clump thickness | 1-10 |
| 3. Uniformity of cell size | 1-10 |
| 4. Uniformity of cell shape | 1-10 |
| 5. Marginal adhesion | 1-10 |
| 6. Single epithelial cell size | 1-10 |
| 7. Bare nuclei | 1-10 |
| 8. Bland chromatin | 1-10 |
| 9. Normal nucleoli | 1-10 |
| 10. Mitosis | 1-10 |
| 11. Class | ( 2 for benign, 4 for malignant) |

The experiment is conducted with this dataset by considering the single and multi layer neural network models. Back propogation algorithm with momentum and variable learning rate is used to train the networks. The experimental
results shown in Table 3 proved that neural networks technique provides satisfactory results for the classification task.

| Training Samples | Test Samples | Classification Efficiency | |
|---|---|---|---|
| | | Single Layer | Multi Layer |
| 50 | 300 | 75% | 80% |
| 100 | 200 | 78.6% | 82% |
| 200 | 100 | 84.2% | 89.4% |
| 300 | 50 | 88.9% | 92% |

Table 3 Experimental Results of Cancer Dataset

## 4.3 NAÏVE BAYES CLASSIFIER

In [4] the Authors Abdelghani Bellaachia & Erhan Guven have performed an analysis of the prediction of survivability rate of breast cancer patients using three data mining techniques the Naïve Bayes, the back-propagated neural network, and C4.5 decision tree algorithms using the Weka toolkit . The Weka is an collection of tools for various data mining techniques like classification, regression, clustering, association rules, and visualization. The toolkit is developed in Java and is an open source software. A newer version of **SEER** database (period of 1973-2002 with 482,052 records) have been used with two additional fields Vital Status Recode (VSR) and the Cause of Death (COD).In this study ,the accuracy of three data mining techniques is compared & experimental results of their approach is shown in table 4.

| Classification Technique | Accuracy(%) |
|---|---|
| Naïve Bayes | 84.5 |
| Artificial Neural Net Neural Net | 86.5 |
| C4.5 | 86.7 |

Table 4.Accuracy of Cancer Dataset

The study shows that the preliminary results are promising for the application of the data mining methods into the survivability prediction problem in medical databases. The achieved prediction performances are comparable to existing techniques. However, C4.5 algorithm has a much better performance than the other two techniques.

## 4.4 SUPPORT VECTOR MACHINES

In [5] the authors have analyzed whether chemotherapy could prolong survival time of breast cancer patients using data mining technique. The principal objective of this work is to try to identify breast cancer patients for whom chemotherapy prolongs survival time. The study is performed on 253 breast cancer Patients with6-feature space consisting of 5 cytological features (mean of area, standard error of area, worst area, worst texture and worst of perimeter) and one pathology feature (tumor size) tumor size.. Three nonlinear smooth support vector machines (SSVMs) is used for classifying breast cancer patients into the three prognostic groups i e Good, Poor and Intermediate. These results suggest that the patients in the Good group should not receive chemotherapy while those in the Intermediate group should receive chemotherapy based on our survival curve analysis.

## 4.5 LOGISTIC REGRESSION

The prediction of breast cancer survivability is a challenging task. In [6] the authors have performed a comparative study of multiple prediction models for breast cancer survivability using a large dataset along with a 10-fold cross-validation. Three different classification models: artificial neural networks, decision trees, and logistic regression have been used in the experiment. In order to perform the research data contained in the SEER Cancer Incidence Public-Use Database for the years 1973—2000 is used. The SEER Breast cancer data consisted of 433,272 records/cases and 72 variables. These 72 variables provide socio-demographic and cancer specific information concerning an incidence of cancer. Each record represents a particular patient—tumor pair within a registry. After using these data cleansing and data preparation strategies, the final dataset, which consisted of 17 variables (16 predictor variables and 1dependent variable) and 202,932 records, was constructed. The dependent variable is a binary categorical variable with two categories: 0 and 1, where 0 denoting did not survive and 1 denoting survived.

Logistic regression is a generalized form of linear regression. It is used primarily for predicting the binary or multi-class dependent variables. As the response variable is discrete, it cannot be modeled directly by linear regression. Therefore, rather than predicting point estimate of the event itself, it builds the model to predict the odds of the occurrence. In a two-class problem, odds greater than 50% would mean that the case is assigned to the class designated as ''1'' and ''0'' otherwise. While logistic regression is a very powerful modeling tool, it assumes that the response variable (the log odds, not the event itself) is linear with respect to the predictor variables.

The models were evaluated based on the accuracy measures –classification accuracy, sensitivity and specificity. The experiments have been performed using 10 fold cross –Validation for each model, and results are reported  based on the average results obtained from the test dataset (the $10^{th}$ fold) for each fold. In comparison to the above studies, we found that the ANN model achieved a classification accuracy of 0.9121 with a sensitivity of 0.9437 and a specificity of 0.8748. The logistic regression model achieved a classification accuracy of 0.8920 with a sensitivity of 0.9017 and a specificity of 0.8786. However, the decision tree (C5) preformed the best of the three models evaluated. The decision tree (C5) achieved a classification accuracy of 0.9362 with a sensitivity of 0.9602 and a specificity of 0.9066.

## 4.6 Bayesian Networks

Bayesian networks are very attractive for medical diagnostic systems because as they can be applied to make inferences in cases where the input data is incomplete.

A Bayesian network (also referred to as *Bayesian belief network*, *belief network*, *probabilistic network*, or *causal network*) consists of a qualitative part, encoding existence of probabilistic influences among a domain's variables in a directed graph, and a quantitative part, encoding the joint probability distribution over these variables. Each node of the graph represents a random variable and each arc represents a direct dependence between two variables. The directed graph is a representation of a factorization of the joint probability distribution. As there can be many graphs that are capable of encoding the same joint probability distribution. In [20] The authors have implemented the Bayesian Belief Network for an automated breast cancer detection support tool. For the application of computer-aided detection in mammography, an interface is designed for the radiologists who can interact with project's Bayesian network learning algorithm. In [21] the authors evaluated three methods for integrating clinical and microarray data and used them to classify publicly available data on breast cancer patients into a poor and a good prognosis group. The focus is on the prediction of the prognosis in lymph node negative breast cancer (without apparent tumor cells in local lymph nodes at diagnosis).The outcome is defined as a variable that can have two values: poor prognosis or good prognosis. Poor prognosis is corresponding to recurrence within 5 years after diagnosis and good prognosis is corresponding to a disease free interval of at least 5 years. If these two groups can be distinguished, patients will be treated more optimally thus eliminating over- or under-treatment.

## 5. CONCLUSION AND FUTURE WORK

To aid clinicians in the diagnosis of breast cancer, recent research has looked into the development of computer aided diagnostic tools. Various data mining techniques have been widely used for breast cancer diagnosis. In this paper we have discuss some of effective techniques that can be used for breast cancer classification. Among the various data mining classifiers and soft computing approaches, Decision tree is found to be best predictor with 93.62% Accuracy on benchmark dataset (UCI machine learning dataset) and also on SEER dataset. In future the predictor can be used to design a web based application to accept the predictor variables and   Automated system Decision Tree based prediction can be implemented in remote areas like rural regions or country sides, to imitate like human diagnostic expertise for prediction of ailment.. The Bayesian network is also found to be a popular technique in medical prediction Particular it has been successfully utilized for Brest cancer prognosis and diagnosis. In future we intend to design and implement such system for web based applications.

# REFERENCES

[1]   Orlando Anunciac¸ ˜ao and Bruno C. Gomes and Susana Vinga and Jorge Gaspar and Arlindo L. Oliveira and Jos´e Rueff ,  A Data Mining Approach for the detection of High-Risk Breast Cancer Groups

[2 ]   Maria-Luiza Antonie, Osmar R. Za¨ıane, Alexandru Coma, .Application of Data Mining Techniques for Medical   Image Classification.Proceeding of second International worshop on Mutimedia data mining(MDM/KDD'2001),in     conjuction     with     ACM     SIGKDD     conference.SAN FRANCISCO,USA,AUG 26,2001.

[3]   Dr. K. Usha Rani,   Parallel Approach for Diagnosis of Breast Cancer using Neural Network Technique.

[4]   Abdelghani   Bellaachia,Erhan   guven,   Predicting   Breast   cancer   survivability   using   Data MiningTechniques.

[5]   Y.-J. Lee _, O. L. Mangasarian y& W. H. Wolberg, Survival-Time Classifcation of Breast Cancer Patients

[6]   Dursun Delen*, Glenn Walker, Amit Kadam, Predicting breast cancer survivability: a comparison of three data mining methods Artificial Intelligence in Medicine (2004)

[7]   Breast cancer Q&A/facts and statistics (http://www.komen. org/bci/bhealth/QA/q_and_a.asp).

[8]   Jerez-Aragone´s JM, Gomez-Ruiz JA, Ramos-Jimenez G, Munoz-Perez J, Alba-Conejo E. A combined neural network and decision trees model for prognosis of breast cancer relapse. Artif Intell Med 2003;27:45—63.

[9]   Sarvestan Soltani A. , Safavi A. A., Parandeh M. N. and Salehi M., "Predicting Breast Cancer Survivability using data mining techniques," Software Technology and Engineering (ICSTE), 2nd International Conference, 2010, vol.2,pp.227-231.

[10] Delen Dursun , Walker Glenn and Kadam Amit , "Predicting breast cancer survivability: a comparison of three data mining methods," Artificial Intelligence in Medicine ,vol. 34, pp. 113-127 , June 2005.

[11] Anunciacao Orlando, Gomes C. Bruno, Vinga Susana, Gaspar Jorge, Oliveira L. Arlindo and Rueff Jose, "A Data Mining approach for detection of high-risk Breast Cancer groups," Advances in Soft Computing, vol. 74, pp. 43-51,2010.

[12] Padmavati J., "A Comparative study on Breast Cancer Prediction Using RBF and MLP," International Journal of Scientific & Engineering Research, vol. 2, Jan. 2011.

[13] Sudhir D., Ghatol Ashok A., Pande Amol P., "Neural Network aided Breast Cancer Detection and Diagnosis",7th WSEAS International Conference on Neural Networks, 2006.

[14] Han J. and Kamber M., Data Mining: Concepts and Techniques, 2nd ed., San Francisco, Morgan Kauffmann Publishers,2001.

[15] Choi J.P., Han T.H. and Park R.W., " A Hybrid Bayesian Network Model for Predicting Breast Cancer Prognosis", J Korean Soc Med Inform, 2009, pp. 49-57.

[16] Shelly Gupta, Dharminder Kumar, Anand Sharma, Data Mining Classification Techniques Applied For Breast Cancer Diagnosis And Prognosis, Issn : 0976-5166 Vol. 2 No. 2 Apr-May 2011 188, Indian Journal Of Computer Science And Engineering.

[17] Dursun Delen, Analysis of cancer data: a data mining approach, The Journal of Knowledge Engineering, Expert Systems, February 2009, Vol. 26, No. 1.

[18] Allan Tucker , Veronica Vinciotti, Xiaohui Liu , David Garway-Heath,A spatio-temporal Bayesian network classifier for understanding visual field deterioration, Artificial Intelligence in Medicine Volume 34 Issue 2, June, 2005.

[19] Lior Rokach and Oded Maimon Top-Down Induction of Decision Trees Classifiers – A Survey , IEEE TRANSACTIONS ON SYSTEMS, MAN AND CYBERNETICS: PART C, VOL. 1, NO. 11, NOVEMBER 2002.

[20] Jyotirmay Gadewadikar , Ognjen Kuljaca1, Kwabena Agyepong, Erol Sarigul3, Yufeng Zheng and Ping Zhang, Exploring Bayesian networks for medical decision support in breast cancer detection, African Journal of Mathematics and Computer Science Research Vol. 3(10), pp. 225-231, October 2010.

[21] Olivier Gevaert1,_, Frank De Smet1,2, Dirk Timmerman3, Yves Moreau1 and Bart De Moor1 Predicting the prognosis of breast cancer by integrating clinical and microarray data with Bayesian networks, Vol. 22 no. 14 2006, pages e184–e190 BIOINFORMATICS.

[22] Xiao-Hui Wang, Bin Zheng, Walter F. Good , Jill L. King, Yuan-Hsiang Chang, Computer-assisted diagnosis of breast cancer using a data-driven Bayesian belief network, International Journal of Medical Informatics 54 (1999) 115–126,

[23] Lucas PJF: Gaag van der LC, Abu-Hanna A. Bayesian networks in biomedicine and health-care. Artif Intell Med 2004, 30:201–214. Survey paper on recent advances in Bayesian networks in health care.

[24] http://www.scribd.com/doc/28249613/Data-Mining-Tutorial

## Authors

Mrs. Shweta Kharya is a Sr. Assistant Professor in Department of Computer Applications at Bhilai Institute of Technology, Durg (C.G.), India. She is a post-graduate from Pt. Ravi Shankar Shukla University, India. She is a Life fellow member of Indian Society for Technical Education. She has the total teaching experience of 6 yrs.
.

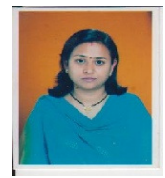